\documentclass[11pt,twoside]{article}
\usepackage{asp2010}

\resetcounters

\begin{document}

\title{Counter-Rotation in Disk Galaxies}

\author{E. M. Corsini}

\affil{Dipartimento di Fisica ed Astronomia `Galileo Galilei',
  Universit\`a di Padova, Padova, Italy}

\begin{abstract}
Counter-rotating galaxies host two components rotating in opposite
directions with respect to each other.  The kinematic and
morphological properties of lenticulars and spirals hosting
counter-rotating components are reviewed. Statistics of the
counter-rotating galaxies and analysis of their stellar populations
provide constraints on the formation scenarios which include both
environmental and internal processes.
\end{abstract}

\vspace{-0.8cm}
\section{Introduction}

Counter-rotating galaxies belong to the class of multi-spin
galaxies. They are characterized by the presence of two components
that are observed rotating in opposite directions with respect to each
other.
Before their discovery, counter-rotating galaxies were considered from
a theorethical point of view and dismissed as elegant curiosities
\citep[see][for an historical perspective on the investigation of
  galactic kinematics]{Rubin1994b}. This belief changed when
\citet{Galletta1987} measured the kinematics of the ionized-gas and
stellar components of the early-type barred galaxy NGC~4564 and showed
they are rotating in opposite directions around the same rotation
axis. In the same period, the first counter-rotating elliptical
\citep[NGC~5898;][]{Bettoni1984, Bertola+Bettoni1988} was found in
Padua too. As more and more data became available, the presence of
counter-rotating components was detected in tens of galaxies along all
the Hubble sequence, from ellipticals to irregulars.

Previous reviews about counter-rotation are those by
\citet{Galletta1996} and \citet{Bertola+Corsini1999}, while
\citet{Corsini+Bertola1998} listed all the counter-rotating galaxies
known at the time.  This paper focuses on the counter-rotation in
lenticulars and spirals.

\section{Varieties of the Counter-Rotating Components}
\label{sec:types}

The phenomenon of counter-rotation is:

\begin{itemize}

\item {\em intrinsic\/} when the two kinematically decoupled
  components are rotating in opposite directions around the same
  rotation axis and, therefore, their vectors of angular momentum are
  antiparallel;

\item {\em apparent\/} if the two kinematically decoupled components
  rotate around skewed rotation axes and the line of sight lies in
  between them so that the vectors of angular momentum are projected
  antiparallel onto the sky plane.

\end{itemize}

\noindent
Observationally, the intrinsic or apparent nature of counter-rotation
may be addressed in not edge-on galaxies by analyzing their full
velocity field as mapped with multi-slit or integral-field
spe\-ctro\-sco\-py.

As far as the counter-rotating components are concerned,
counter-rotation occurs in a variety of forms:

\begin{itemize}

\item {\em gas-versus-stars counter-rotation\/} (also known as {\em
  gaseous counter-rotation\/}) is observed when the gaseous disk
  counter-rotates with respect to the stellar body of the galaxy. This
  is the case of the SB0/SBa NGC~4546 in which the rotation of the
  ionized, molecular, and atomic gas ($\approx10^8$\,M$_{\odot}$) has
  similar amplitude but opposite direction with respect to the stars
  \citep{Galletta1987, Bettoni+1991, Sage+Galletta1994}.

\item {\em stars-versus-stars counter-rotation\/} (also known as {\em
  stellar counter-rotation\/}) occurs when two stellar components
  counter-rotate. Usually the more massive component is labeled as the
  prograde one. The E7/S0 NGC~4550 hosts two cospatial
  counter-rotating stellar disks, one of them is corotating with the
  gaseous disk \citep{Rubin+1992, Johnston+2013, Coccato+2013}. The
  two stellar disks have similar luminosities, sizes, and masses
  \citep{Rix+1992} although one is thicker than the other
  \citep{Cappellari+2007}. The bulge \citep[e.g.,
    NGC~524;][]{Katkov+2011}, a secondary bar \citep[e.g.,
    NGC~2950;][]{Corsini+2003a, Maciejewski2006}, or some of the stars
  in a bar \citep{Bettoni1989, Bettoni+Galletta1997} are other
  examples of counter-rotating stellar components.

\item {\em gas-versus-gas counter-rotation\/} is reported when two
  gaseous disks counter-rotate. The S0 NGC~7332 possesses two apposed
  disks of ionized gas rotating in opposite sense with respect to each
  other. The gaseous material ($\approx10^5$\,M$_\odot$) displays
  non-circular motions indicating it has not reached equilibrium
  \citep{Fisher+1994, Plana+Boulesteix1996}.

\end{itemize}

\begin{figure*}[!t]
\centering
\includegraphics[width=0.45\textwidth, 
  bb=70 100 540 740, clip=]{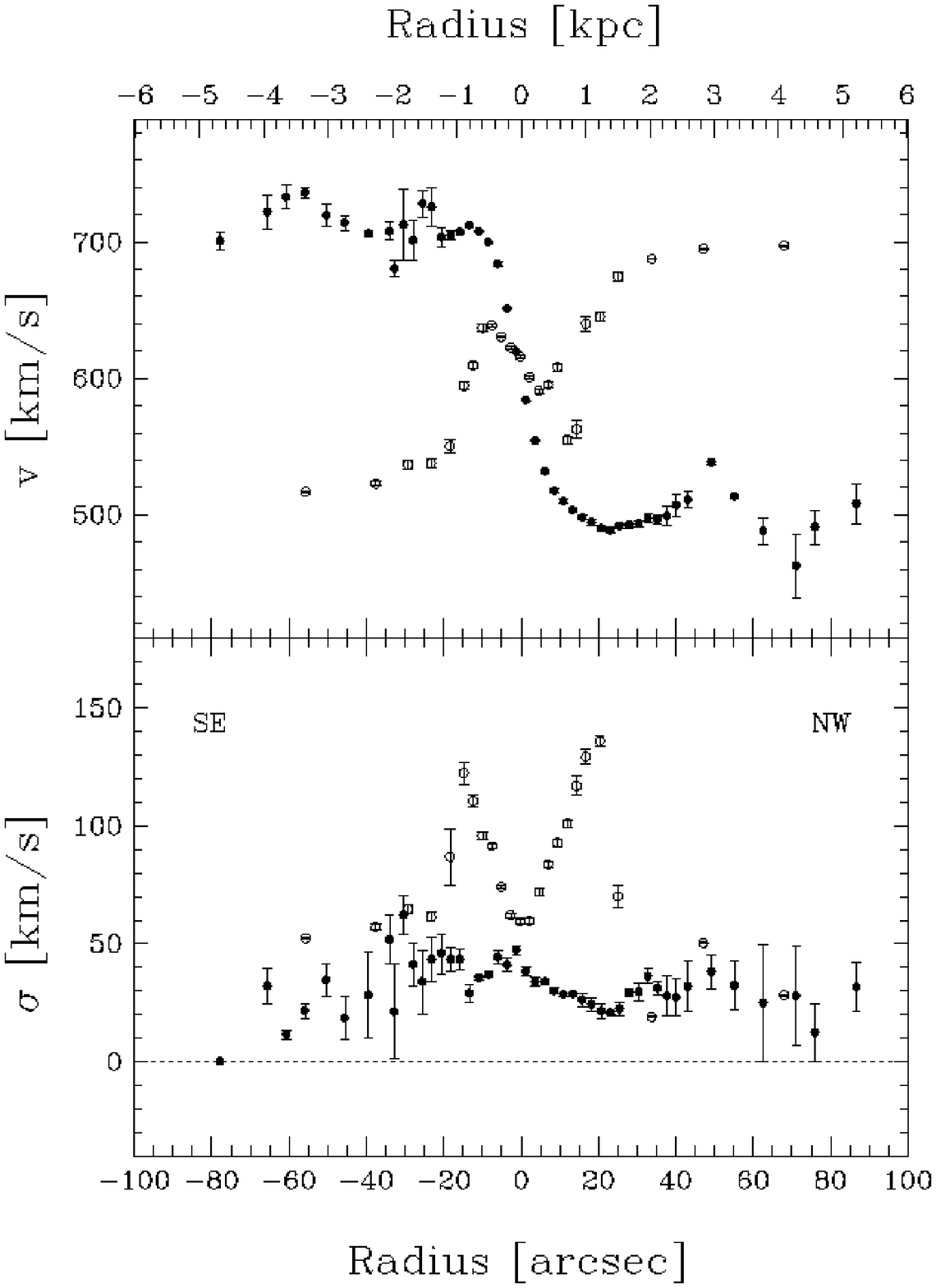}
\hspace{0.3cm}
\includegraphics[width=0.45\textwidth, 
  bb=70 100 540 740, clip=]{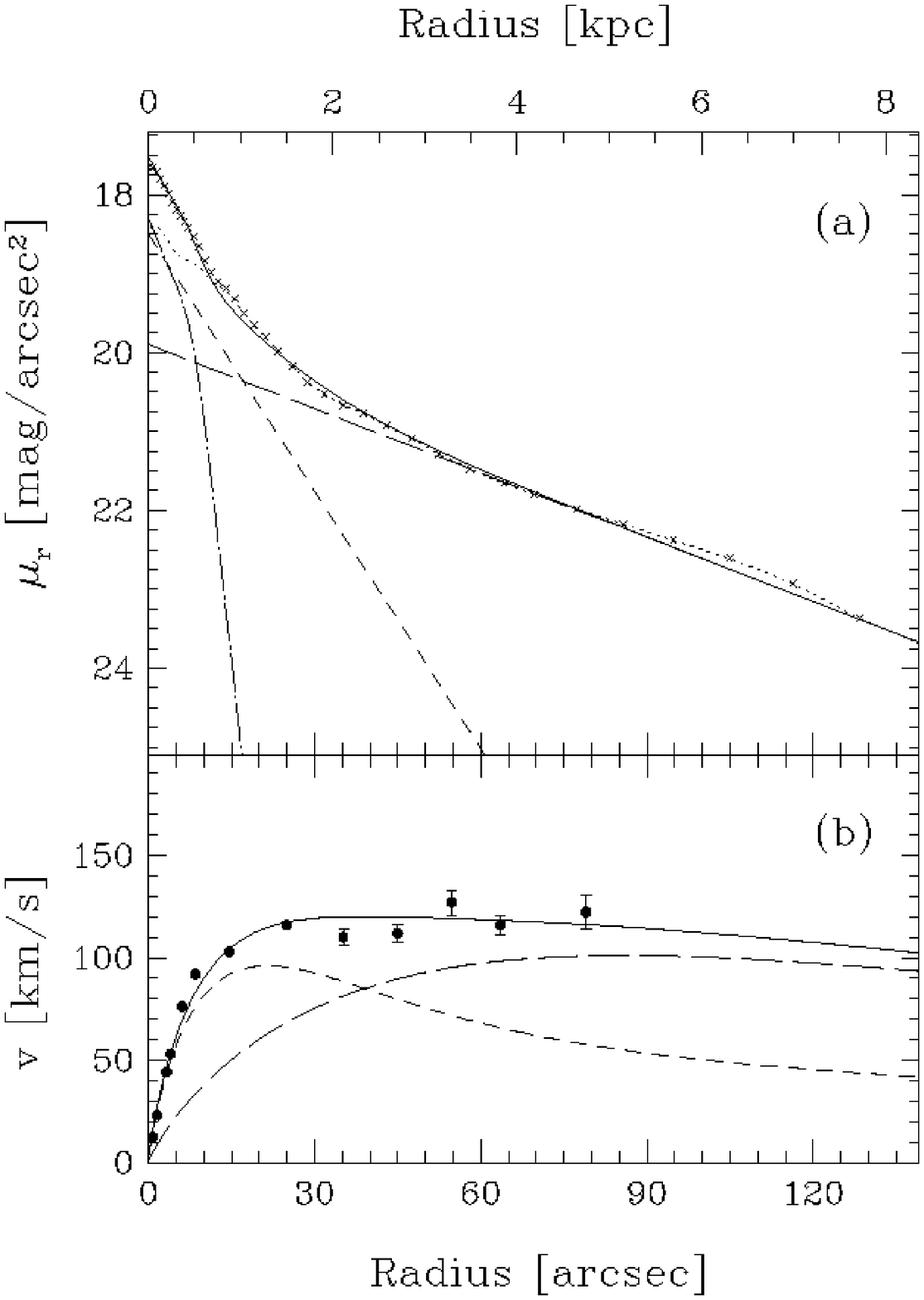}
\caption{The inner stellar counter-rotation and overall gaseous
  counter-rotation of the Sa NGC~3593.  Left panels: Velocity (top
  panel) and velocity dispersion (bottom panel) radial profiles
  measured along the major axis of NGC~3593 for the stellar (open
  circles) and ionized-gas component (filled circles).  Right panels:
  Photometric decomposition of NGC~3593 (top panel). The
  surface-brightness radial profile measured along the major axis
  (crosses) is decomposed into the contribution of a bulge (dot-dashed
  line), a smaller-scale disk (short-dashed line), and a larger-scale
  disk (long-dashed line). The sum of the three component is given by
  the continuous line.  Mass model of NGC~3593 (bottom panel). The
  contribution of the smaller-scale disk (short-dashed line) and
  larger-scale disk (long-dashed line) to the total circular velocity
  (continuous line) is shown with the observed ionized-gas velocity
  curve (filled circles).  The contribution of the bulge is
  neglected. From \citet{Bertola+1996}.}
\label{fig:n3593}
\end{figure*}

Finally, counter-rotation in disk galaxies is detected in:

\begin{itemize}

\item the {\em inner regions\/} of the galaxy. For example, the Sa
  NGC~3593 (Fig.~\ref{fig:n3593}) is composed by a small bulge, a main
  stellar disk which contains $\sim80\%$ of the stars
  ($1.2\,\times\,10^{10}$\,M$_{\odot}$) and a secondary
  counter-rotating stellar disk \citep{Bertola+1996,
    Coccato+2013}. The latter dominates the kinematics in the inner
  kpc and corotates with the disk of ionized and molecular gas
  \citep{Corsini+1998, GarciaBurillo+2000}.

\item the {\em outer regions\/} of the galaxy. The Sab NGC~4826 (M64)
  contains two counter-rotating nested disks of ionized, molecular
  and neutral gas extending out $\sim1$ and $\sim11$ kpc, respectively
  \citep{Braun+1992, Braun+1994, Rubin1994a, Walterbos+1994}. They
  have similar masses ($\approx10^8$\,M$_\odot$) and are both coplanar
  to the stellar disk. Stars corotate with the inner gas. Beyond the
  dust lane which marks transition between the two gaseous disks, only
  a small fraction of stars ($\lesssim5\%$) corotate with the outer
  gas \citep{Rix+1995}.

\item {\em overall the galaxy,\/} as for the Sa NGC~3626. This is the
  first spiral galaxy where the gaseous component
  ($\approx10^9$\,M$_\odot$) was observed to counter-rotate at all
  radii with respect to the stars \citep{Ciri+1995,
    GarciaBurillo+1998, Haynes+2000, Silchenko+2010}.

\end{itemize}

\section{Detection of Counter-Rotation}
\label{sec:kinematics}

The detection of a counter-rotating gaseous component is usually
straightforward. Observationally, it may be addressed by looking at
the opposite orientation of the ionized-gas emission lines and stellar
absorption lines in 2-dimensional optical spectra \citep[e.g., see
  Fig.~1 in][]{Galletta1987} or in position-velocity diagrams
\citep[e.g., see Fig.~1 in][]{Bureau+Chung2006}. Moreover, the
standard techniques adopted to derive the kinematics of gas and stars
allow to measure differences of few km\,s$^{-1}$ in their rotation
velocities.

On the contrary, unveiling a counter-rotating stellar component is a
more difficult task. It requires a detailed data analysis because the
kinematics of two counter-rotating stellar populations are measured
from the same absorption lines. X-shaped absorption lines are only
observed when the two components are photometrically similar
\citep[see Figs.~2c and 2d in][]{Rubin+1992}. However, a bimodal
line-of-sight velocity distribution (LOSVD) is the signature of the
presence of two counter-rotating components \citep[see Fig.~2
  in][]{Rix+1992}. But, the detection of the LOSVD bimodality depends
on both the galaxy properties (i.e., the fraction, dynamical status,
and velocity of the retrograde stars with respect to the prograde
ones) and instrumental setup (i.e., spectral sampling and resolution)
of the spectroscopic observations.
By analyzing synthetic spectra with a different fraction of
counter-rotating stars, \citet{Kuijken+1996} and \citet{Pizzella+2004}
set an upper limit of $\sim10\%$ on the fraction of retrograde stars
which can be detected in long-slit spectra of intermediate resolution
($\sigma_{\rm inst}\,\simeq\,50$ km\,s$^{-1}$) obtained with a spatial
resolution of $\rm FWHM\,\simeq\,1''$ and a signal-to-noise ratio
$S/N\,\geq\,30$ \AA$^{-1}$. Similar results were obtained by
\citet{Coccato+2013} for integral-field spectra.

The LOSVD is poorly reproduced by a Gauss-Hermite expansion
\citep{Gerhard1993, vdMarel+Franx1993} when the galaxy hosts a
secondary kinematic component \citep{Fabricius+2012, Katkov+2013}. In
addition, noise and aliasing features in the LOSVD can mimic what may
be interpreted as a counter-rotating component. Therefore, the
recovery and decomposition of parametric LOSVDs has to be performed
with caution. Indeed, the counter-rotating bulge of the Sb NGC~7331
found by \citet{Prada+1996} was proved by \citet{Bottema1999} to be an
artifact of the method adopted to measure the stellar kinematics. This
seems also the case of the counter-rotating stellar disks of the Sb
spiral NGC~7217. Fabricius et al. (in prep., but see also this volume)
show that the stellar components of NGC 7217 are corotating in spite
of what previously claimed by \citet{Merrifield+Kuijken1994}.

There is compelling evidence that the presence of two off-center and
symmetric peaks in the stellar velocity dispersion in combination with
zero velocity rotation measured along the galaxy major axis is
indicative of two counter-rotating disks. This kinematic features are
observed in the radial range where the two counter-rotating components
have roughly the same luminosity and their LOSVDs are unresolved
\citep[][see Fig.~\ref{fig:n3593}, left panels]{Bertola+1996}.
Recently, \citet{Krajnovic+2011} have found 11 galaxies (including
NGC~4550) with a double-peaked velocity dispersion in the
volume-limited sample of 260 nearby early-type galaxies gathered by
the ATLAS-3D project. They are interesting candidates for a further
investigation to address the fraction of their counter-rotating stars.

\section{Environment and Morphology of Counter-Rotating Galaxies}
\label{sec:morphology}

The morphology of most of the galaxies hosting counter-rotating
components appear undisturbed with no evidence of recent interaction
with small satellites or companions of similar size. Indeed, the
environment of counter-rotating galaxies does not appear statistically
different from that of normal galaxies, as pointed out by
\citet{Bettoni+2001}. They investigated the number, size, and
distribution of the faint satellites (to a limiting magnitude
$B<21.5$) and bright companions (within a searching radius $R\,<\,0.6$
Mpc and redshift difference $\Delta V\,<\,600$ km\,s$^{-1}$) and the
large-scale environment of 49 galaxies with counter-rotation and 43
comparison galaxies without counter-rotation.

These findings set constraints on the origin of counter-rotation,
because the formation process is required to not affect the present
morphology of the host galaxies and the galaxy density of their
surrounding regions. Therefore, retrograde gas accretion has to be a
smooth and non-traumatic process, major mergers are expected to occur
only early in the life of the host galaxy whereas minor mergers may be
more recent.
However, the relics of merger events like the collisional debris and
tidal tails are generally transient and faint structures. Their
surviving ages vary from a few hundred Myr to a few Gyr and they have
a surface brightness that is typically $25$ $B$-mag arcsec$^{-2}$ when
young and below 27 $B$-mag arcsec$^{-2}$ when getting older. The
detection of these fine structures requires deep optical imaging, but
the comparison of their morphology and kinematics with the results of
numerical simulations promises to constrain the epoch and mechanism of
the second event \citep[e.g.,][]{Corsini+2002, Duc+2011}

As far as the morphology of the host galaxy concerns, no
counter-rotating components are detected in late-type spiral
galaxies. Three of the few spirals hosting counter-rotating gaseous
and/or stellar disks (NGC~3593, NGC~3626, and NGC~4138) belong to the
same morphological type, being very early-type spirals (S0/a--Sa) with
smooth arms. Their spiral pattern is either defined entirely or
dominated by the dust lanes. Indeed, they appear in same section of
the The Carnegie Atlas of Galaxies \citep[Plates
  72-76;][]{Sandage+Bedke1994}.
The suppression of arms in counter-rotating spirals has been recently
recognized in high-resolution N-body simulations of multi-armed spiral
features triggered through swing amplification by density
inhomogeneities (with the mass and lifetime of the order of a typical
giant molecular cloud) orbiting the disk \citep{DOnghia+2013}.
A survey of a sample of early S0/a and Sa spirals selected to have the
spiral pattern traced by dust lanes unveiled the presence of
kinematically decoupled gas components but no new case of
counter-rotation \citep{Corsini+2003b}.

Previous 2-dimensional N-body simulations of disk galaxies with a
significant fraction of counter-rotating stars predicted the formation
of a stationary and persisting one-arm leading spiral wave (with
respect to the corotating stars) due to the two-stream disk
instability \citep{Lovelace+1997, Comins+1997}. However, the
counter-rotating spirals studied so far have intermediate-to-high
inclination which makes difficult to identify the presence of a
one-armed spiral pattern, whereas kinematic data for low inclined
one-arm systems are missing.

\section{Statistics of Counter-Rotation}
\label{sec:statistics}

By analyzing the existing data of S0 galaxies for which the kinematics
of the ionized gas and stellar components were measured along (at
least) the major axis, \citet{Pizzella+2004} found a counter-rotating
(or a kinematically decoupled) gaseous component in 17/53
galaxies. This fraction corresponds to $32\%^{+19}_{-11}$ (at the 95\%
confidence level) and it is consistent with previous statistics by
\citet[][$35\%$]{Bertola+1992}, \citet[][$(24\pm8)\%$]{Kuijken+1996},
and \citet[][$24\%^{+8}_{-6}$]{Kannappan+Fabricant2001}. These
findings favor a scenario in which acquisition events in S0 galaxies
are not limited to few peculiar objects but they are a widespread
phenomenon \citep{Bertola+1992}, as it is for ellipticals
\citep{Bertola+1988}.
The recent results by \citet{Davis+2011} based on integral-field
spectroscopy and radio observations support this scenario too. They
found that the gas in $(36\pm5)\%$ (40/111) of their sample of
fast-rotating early-type galaxies is kinematically misaligned with
respect to the stars. In addition, the ionized, molecular, and atomic
gas in all the detected galaxies are always kinematically aligned,
even when they are kinematically misaligned from the stars. This
implies that all these phases of the interstellar medium share a
common origin and underlines the role of external acquisition.

In contrast to the prevalence of counter-rotating gas in S0 galaxies,
\citet{Kuijken+1996} estimated than $<10\%$ (at the 95\% confidence
level) of S0 galaxies host a significant fraction ($>5\%$) of
counter-rotating stars.

\citet{Pizzella+2004} addressed the frequency of counter-rotation in
spiral galaxies from a sample of 50 S0/a−-Scd galaxies, for which the
major-axis kinematics of the ionized gas and stars were obtained with
the same spatial and spectral resolution, and measured with the same
analysis techniques. It turns out that $<12\%$ and $<8\%$ (at the 95\%
confidence level) of the sample galaxies host a counter-rotating
gaseous and stellar disk, respectively.  For comparison,
\citet{Kannappan+Fabricant2001} set an upper limit of 8\% on the
fraction of spirals hosting counter-rotating gas by analyzing a sample
of 38 Sa--Sbc galaxies.

\section{Formation Scenarios: External and Internal Processes}
\label{sec:formation}

Numerical simulations show that episodic or prolonged accretion of gas
from environment \citep{Thakar+Ryden1996, Thakar+Ryden1998} and
merging with a gas-rich dwarf companion \citep{Thakar+Ryden1996,
Thakar+1997} are viable mechanisms for the retrograde acquisition of
small amounts of external gas. They give rise to counter-rotating
gaseous disks only in S0 galaxies, since in spiral galaxies the
acquired gas is swept away by the pre-existing gas. Therefore, the
formation of counter-rotating gaseous disks is favored in S0 galaxies
since they are gas-poor systems, while spiral disks host large amounts
of gas \citep{Bettoni+2003}, which is corotating with the stellar
component. When gas-rich systems acquire external gas in retrograde
orbits, the gas clouds of the new retrograde and pre-existing prograde
components collide, lose their centrifugal support, and accrete toward
the galaxy center. A counter-rotating gaseous disk will be observed
only if the mass of the newly supplied gas exceeds that of the
pre-existing one \citep{Lovelace+Chou1996}. A counter-rotating stellar
disk is the end-result of star formation in the counter-rotating gas
component. For this reason we observe a larger fraction of
counter-rotating gaseous disks in S0s than in spirals
\citep{Pizzella+2004}. This also explains why the mass of
counter-rotating gas in most S0 galaxies is small compared to that of
the stellar counter-rotating components \citep{Kuijken+1996}.
Counter-rotating gaseous and stellar disks in spirals are both the
result of retrograde acquisition of large amounts of gas, and they are
observed with the same frequency \citep{Pizzella+2004}. In this
framework, stellar counter-rotation is the end result of star
formation in a counter-rotating gaseous disk. The formation of two
counter-rotating stellar disks from material accreted from two
distinct filamentary structures in cosmological simulations has been
recently discussed by \citep{Algorry+2014}.

Usually, major mergers between disk galaxies with comparable masses
are ruled out because they tend to produce ellipticals. Anyway, for a
narrow range of initial conditions, major mergers are successful in
building a remarkably axisymmetric disk which hosts two
counter-rotating stellar components of similar mass and size
\citep[see also Bettoni et al., this
  volume]{Puerari+Pfenniger2001}. Moreover, the coplanar merging of
two counter-rotating progenitors heats more the prograde than the
retrograde stellar disk and the gas ends up aligned with the total
angular momentum (dominated by the orbital angular momentum), and thus
with the prograde stellar disk, as observed in NGC~4550
\citep{Crocker+2009}.

An alternative to the external-origin scenarios has been proposed by
\citet{Evans+Collett1994} for NGC~4550, and it involves the
dissolution of a bar or a triaxial stellar halo. In this process, the
stars moving on box orbits escape from the confining azimuthal
potential well and move onto tube orbits. In non-rotating disks, there
are as many box orbits with clockwise azimuthal motion as with
counter-clockwise. Thus, half box-orbit stars are scattered onto
clockwise-streaming tube orbits, half onto counter-clockwise ones. In
this way, two identical counter-rotating stellar disks can be built.

Since barred galaxies host quasi-circular retrograde orbits, the
origin of the stellar counter-rotation observed in barred galaxies
\citep{Bettoni1989, Bettoni+Galletta1997} can be the result of
internal dynamical processes
\citep{Wozniak+Pfenniger1997}. Nevertheless, accreted gas can be
trapped on this family of retrograde orbits and then eventually form
new stars.
These counter-rotating material may lead to the formation of a
secondary bar rotating in opposite direction with respect to the main
one \citep{Sellwood+Merritt1994}. Observationally,
the formation of secondary bars may be constrained by addressing the
occurrence of counter-rotating secondary bars. Indeed, the most
accepted view on the origin of secondary bars is that they form
through instabilities in gas inflowing along the main bar
\citep{Shlosman+1989}. But, a retrograde bar is unlikely to be
supported by a prograde disk. Numerical simulations suggest that two
counter-rotating nested bars, formed in two counter-rotating stellar
disks that overlap each other, are stable and long-living systems
\citep{Friedli1996}. This leads to the
possibility that secondary bars form out of inner stellar disks, like
those observed in the nuclei of several disk galaxies
\citep{Pizzella+2002, Ledo+2010}. To date counter-rotating nuclear
disks have been detected only in elliptical galaxies
\citep[e.g.,][]{Morelli+2004}.

\section{Stellar Populations of Counter-Rotating Components}
\label{sec:populations}

The different formation mechanisms of counter-rotating disk galaxies
are expected to leave different signatures in the properties of the
prograde and retrograde stellar populations. In particular, their age
difference may be used to discriminate between competing scenarios for
the origin of counter-rotation.

The gas accretion followed by star formation always predicts a younger
age for the counter-rotating component, whereas the counter-rotating
component formed by the retrograde capture of stars through minor or
major mergers may be either younger or older with respect to the
pre-existing stellar disk. The external origin also allows that the
two counter-rotating components have different metallicities and
$\alpha$-enhancements. In contrast, the formation of large-scale
counter-rotating stellar disks due to bar dissolution predicts the
same mass, chemical composition, and age for both the prograde and
retrograde components.

A spectroscopic decomposition that separates the relative contribution
of the coun\-ter-rotating stellar components to the observed galaxy
spectrum is therefore needed to disentangle their stellar populations.
This has been recently done for the counter-rotating stellar disks of
NGC~3593 \citep{Coccato+2013}, NGC~4550 \citep{Coccato+2013,
  Johnston+2013}, and NGC~5719 \citep{Coccato+2011}. In all of them,
the counter-rotating stellar disk rotates in the same direction as the
ionized gas, and it is less massive, younger, more metal poor, and
more $\alpha$-enhanced than the main stellar disk. These findings rule
out an internal origin of the secondary stellar component and favor a
scenario where it formed from gas accreted on retrograde orbits from
the environment fueling an {\em in situ\/} outside-in rapid star
formation.

\begin{figure*}[!t]
\centering \includegraphics[angle=0.0, width=0.8\textwidth, 
  bb=40 270 520 520, clip=]{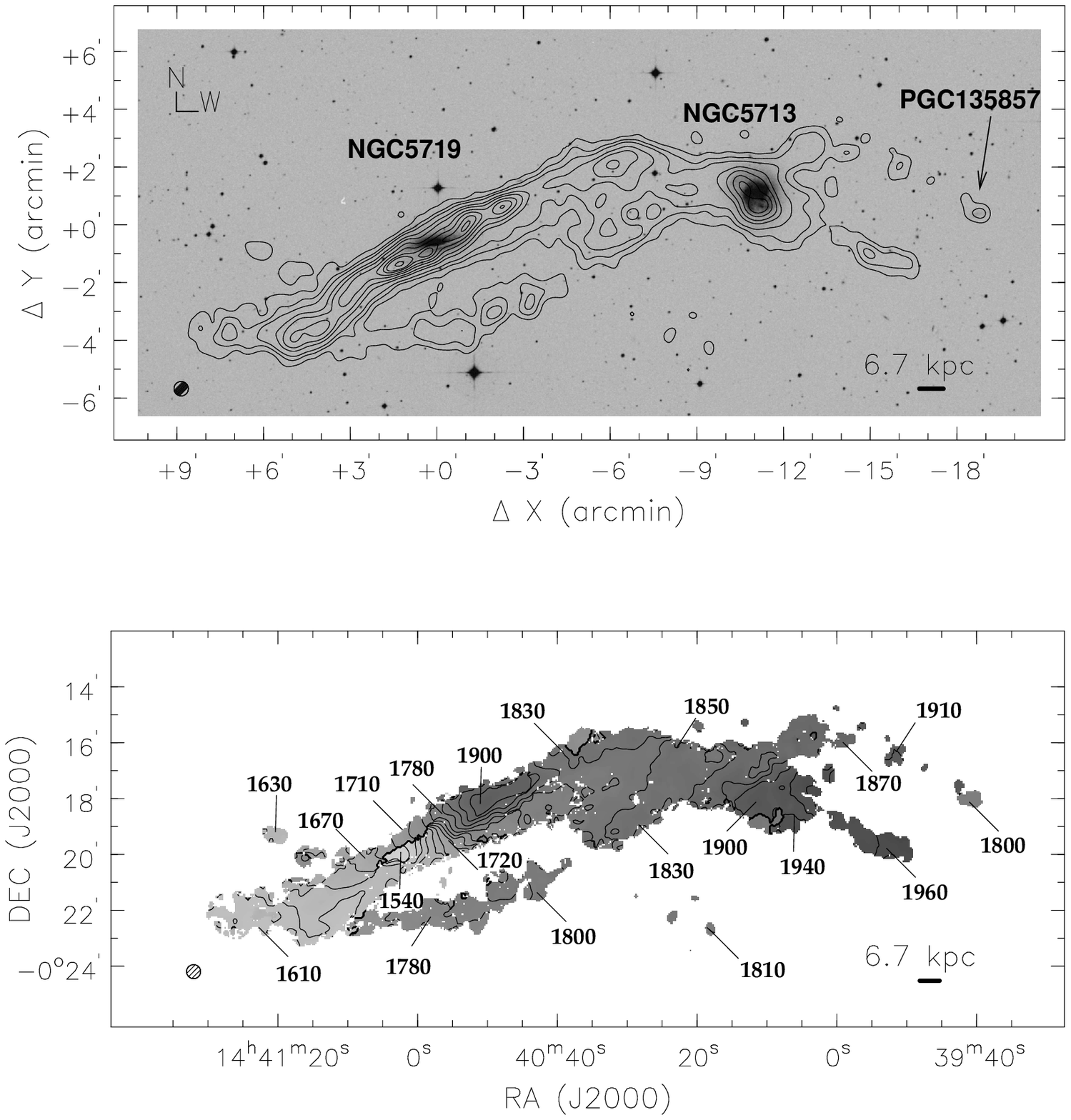}
\centering \includegraphics[angle=0.0, width=0.8\textwidth, 
  bb=30 240 570 620, clip=]{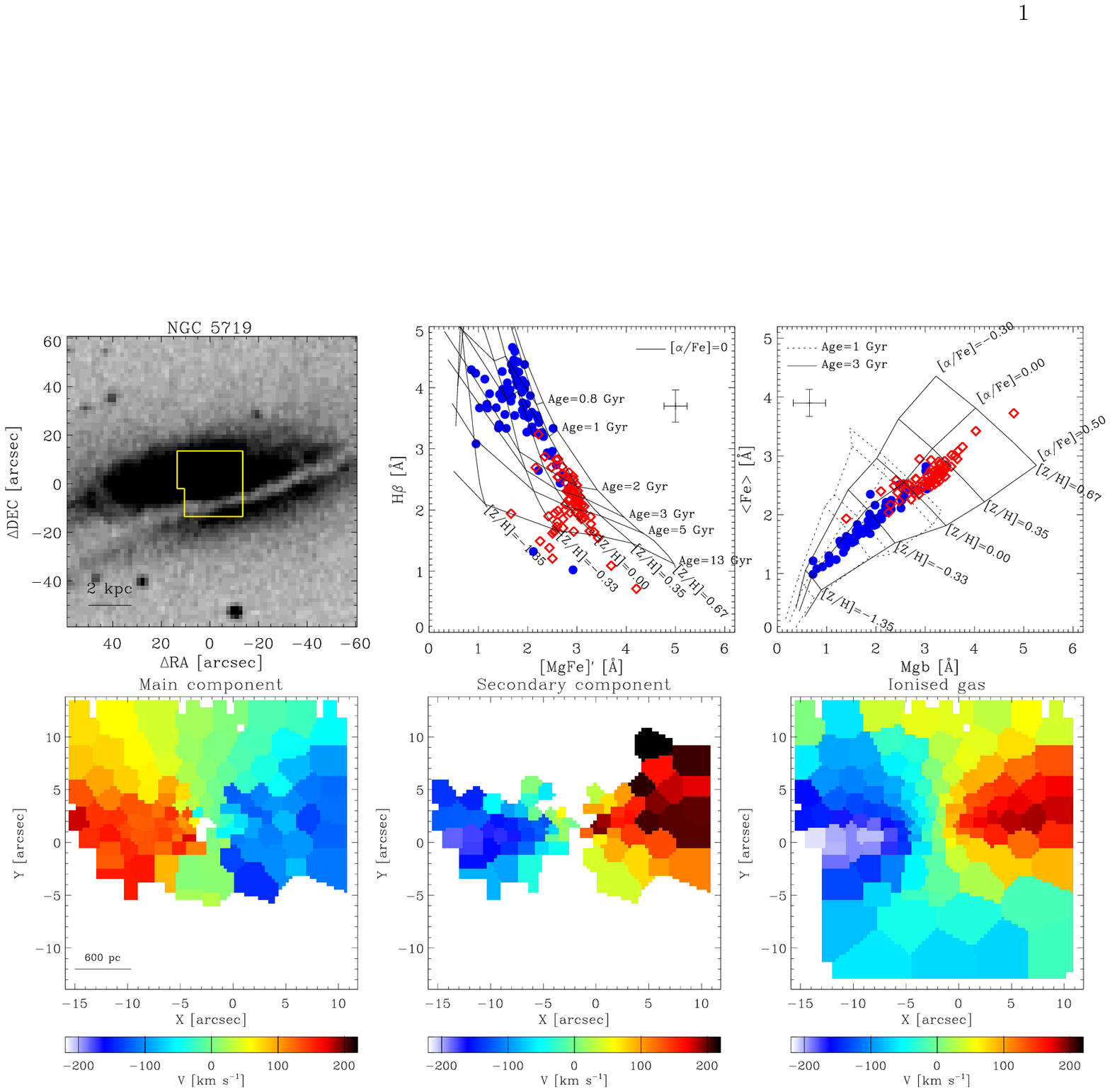}
\caption{The counter-rotating stellar disks of the interacting Sab
  spiral ga\-la\-xy NGC~5179. Top panels: Contour map of the
  \ion{H}{I} column density distribution of NGC~5719 superimposed on
  an optical image from the Digitized Sky Survey. The lowest contour
  level is at $7.0\,\times\,10^{19}$ atoms\,cm$^{-2}$ and the
  increment is $2.4\,\times\,10^{20}$ atoms\,cm$^{-2}$. From
  \citet{Vergani+2007}. Middle panels: Field of view observed with
  integral-field spectroscopy (left panel) and measured equivalent
  width of the line-strength indices H$\beta$ vs. Mg~{\it b} (central
  panel) and $\langle$Fe$\rangle$ vs. [MgFe]' (right panel) with the
  predictions for age, metallicity, and $\alpha$-enhancement from
  single stellar population models.  Each spatial bin returns the
  indices of both the main stellar (red diamonds) and secondary
  counter-rotating stellar component (blue circles). From
  \citet{Coccato+2011}.  Bottom panels: Two-dimensional velocity
  fields of the main stellar component (left panel), secondary stellar
  component (central panel), and ionized-gas component (right
  panel). From \citet{Coccato+2011}.}
\label{fig:n5719}
\end{figure*}

The Sab spiral NGC~5179 shows a spectacular on-going interaction with
its face-on Sbc companion NGC~5713 (Fig.~\ref{fig:n5719}, top
panel). The interaction is traced by a tidal bridge of neutral
hydrogen, which feeds the counter-rotating gaseous and stellar
components \citep{Vergani+2007}. NGC~5719 is the first interacting
disk galaxy in which counter-rotation has been detected
(Fig.~\ref{fig:n5719}, bottom panels). The age of counter-rotating
stellar population ranges from 0.7 to 2.0 Gyr and metallicity changes
from subsolar ($[Z/{\rm H}]\,\simeq\,-1.0$ dex) in the outskirts to
supersolar ($[Z/{\rm H}]\,\simeq\,0.3$ dex) in the center, where\-as
the main stellar component has ages ranging from 2 to 13.5 Gyr and
nearly solar metallicity. The youngest ages and highest metallicities
are found in correspondence of the star forming regions. The
$\alpha$-enhancement of the counter-rotating component indicates a
star formation history with a time-scale of 2 Gyr
(Fig.~\ref{fig:n5719}, middle panels).
On the contrary, the formation through a major galaxy merger cannot be
completely ruled out for NGC~3593 and NGC~4550, which are both quite
isolated and undisturbed galaxies \citep{Coccato+2011, Coccato+2013}.
A larger sample is required to understand by statistical arguments
whether it was accretion or merger the most efficient mechanism to
assembly counter-rotating spirals.

\section{Concluding Remarks}

After they were discovered three decades ago, counter-rotating
galaxies still represent a challenging subject for both theorists and
observers. Although the broad picture of the formation of
counter-rotating galaxies is in place, we still miss many details.

A few issues should be attacked first in the near future to make a
step forward in our understanding of counter-rotation in disk
galaxies. They include: a deep imaging survey to look for the
fingerprints of accretion and merging events in the environment of
counter-rotating galaxies at very low levels of surface brightness;
the analysis of a complete sample of spiral galaxies to drive unbiased
conclusions about the frequency of the different kinds of
counter-rotation; the derivation of the stellar LOSVD from high
(spectral and spatial) resolution data obtained with wide-field
integral-field units to look for yet undetected retrograde stars; and
the extensive measurement of the stellar populations of the prograde
and retrograde components in counter-rotating galaxies to test the
predictions of the different formation scenarios.

\end{document}